\documentclass[12pt]{article}
\usepackage{amsfonts}
\begin{document}
\begin{center}
{\Large\bf (Super)twistors and (super)strings}\\[0.4cm]
{\large D.V. Uvarov}\\[0.4cm]
{\it Kharkov Institute of Physics and Technology,}\\[0.2cm]
{\it 61108 Kharkov, Ukraine}\\
\end{center}
\begin{abstract}
The Lagrangian formulation of the $D=4$ bosonic string and superstring in terms of the (super)twistors is considered. The (super)twistor form of the equations of motion is derived and the $\kappa-$symmetry transformation for the supertwistors is given. It is shown that the covariant $\kappa-$symmetry gauge fixation results in the action quadratic in the (super)twistor variables.
\end{abstract}

\section{Introduction}

Twistor theory \cite{twistor} provides deep insight into the structure and geometry of space-time using spinor methods that led to the series of outstanding results in the field theory and gravitation \cite{WW}. Recently the supersymmetric generalization \cite{Ferber} of twistor theory attracted much attention because of the discovered correspondence \cite{Witten} between some of the perturbative properties of $N=4$ superYang-Mills theory and the string models in certain supertwistor spaces.

In view of these results it seems interesting to investigate the supertwistor formulations of other supersymmetric objects. Such (super)twistor formulations were previously widely explored for various (super)particle \cite{Ferber}, \cite{Shirafuji}-\cite{Bette}, tensionless (super)string and (super) $p$-brane models \cite{Ilyenko}-\cite{bz03} both on the classical and quantum levels. One of the most interesting features characterizing the supertwistor formulations is the exclusion of some of the unphysical pure gauge variables present in the (super)space-time formulations due to the Penrose variables transform without the Lorentz-covariance breaking that was first discussed in Ref.\cite{Shirafuji}. This, in particular, concerns the exclusion of the redundant components of Grassmann coordinates $\theta$ related to $\kappa$-symmetry \cite{DAL}, \cite{Siegel} of the action functional in the superspace formulation. As an extension of the above results it is natural to examine a possibility of the supertwistor reformulation of the tensile Green-Schwarz superstring action \cite{GSW} and to verify whether and under what conditions such pure gauge variables exclusion might take place. Previous attempts \cite{Shaw}-\cite{Zunger} on the incorporation of (super)twistors into the tensile (super)string models\footnote{In \cite{BAPV} there were considered tensile superstring and super $p-$brane models in the superspace with extra bosonic coordinates described by the totally antisymmetric tensors. Corresponding Lagrangians are $\kappa-$invariant without introduction of the Wess-Zumino terms and the transition to the generalized supertwistor variables proceeds in the same way as for the superparticles.} did not shed much light on that. In the present paper we examine this problem choosing as a starting point the Lorentz-harmonic formulation of the tensile (super)string action \cite{BZstring}. That formulation was shown to be classically equivalent to the Green-Schwarz and Polyakov actions but the introduction of the auxiliary Lorentz-harmonic variables\footnote{Harmonic variables were initially introduced in \cite{GIKOS} (see also the monograph \cite{Ivanov}) in connection with the unconstrained superfield formulation of supersymmetric field theories with extended supersymmetry.} \cite{Sokatchev}-\cite{harmonics}, that coincide in the $D=4$ space-time with the normalized Newman-Penrose dyad \cite{twistor}, covariantly resolves the Virasoro constraints resulting in the irreducible realization of the $\kappa$-symmetry transformations\footnote{Note that the Lorentz-harmonic approach of \cite{BZstring} to the formulation of the $N=1$ superstring can be regarded as the component version of the doubly supersymmetric superfield action of the superembedding approach (see \cite{HSW}, \cite{DS} and refs. therein) with all the auxiliary fields except for the Lorentz-harmonics and zweibein being eliminated.}.

In the next section as the preliminary step we consider the twistor formulation of the bosonic string action that in section 3 is generalized to the case of the $D=4$ $N=1$ superstring. We establish that the reduction of the redundant components of Grassmann coordinates is not automatic upon the Penrose variables transform unlike the superparticle and tensionless superstring and super $p$-brane models but requires the utilization of the rheotropic relations equating the projections of the space-time superinvariant Cartan forms to the world-sheet zweibein components that amounts to the implicit $\kappa$-symmetry gauge fixation. Then we turn to the exploration of the $D=4$ $N=2$ superstring and find that for such reduction it is not sufficient even to utilize the rheotropic relations because the Wess-Zumino term includes the nonlinear contribution of the fermionic variables so the desired exclusion of the pure gauge Grassmann variables takes place only upon the explicit (but covariant) $\kappa$-symmetry gauge fixation. The last section contains conclusions and the discussion of further lines of investigation.

\section{Twistor formulation of bosonic string action}

Twistor variables defined as the fundamental spinors of the twistor group $SU(2,2)$ that is the double covering of the conformal group $SO(2,4)$ allow the following decomposition on $SL(2,{\bf C})$ 2-component spinors
\begin{equation}
Z^a=(\mu^\alpha,\ \bar u_{\dot\alpha}),\quad W^a=(\nu^\alpha,\ \bar v_{\dot\alpha}),\quad a=1,...,4;\ \alpha,\dot\alpha=0,1,
\end{equation}
where $\mu^\alpha$ and $\nu^\alpha$ are called the main parts of the twistors $Z^a$ and $W^a$, while $\bar u_{\dot\alpha}$, $\bar v_{\dot\alpha}$ constitute the projectional parts. The conjugation reverses the chirality yielding the dual twistors
\begin{equation}
\bar Z_a=(u_\alpha,\ \bar\mu^{\dot\alpha}),\quad\bar W_a=(v_\alpha,\ \bar\nu^{\dot\alpha})
\end{equation}
but unlike the $SL(2,{\bf C})$ case there are no analogues of the unit tensors $\varepsilon^{\alpha\beta}$, $\varepsilon_{\alpha\beta}$ and $\varepsilon^{\dot\alpha\dot\beta}$, $\varepsilon_{\dot\alpha\dot\beta}$ to raise and lower indices of twistors.

For the introduced twistors to be related to the points of the real $D=4$ Minkowski space-time there should be imposed the zero norm conditions
\begin{equation}\label{3}
Z^a\bar Z_a\equiv\mu^\alpha u_\alpha+\bar u_{\dot\alpha}\bar\mu^{\dot\alpha}=0,\quad W^a\bar W_a\equiv\nu^\alpha v_\alpha+\bar v_{\dot\alpha}\bar\nu^{\dot\alpha}=0
\end{equation}
and the incidence conditions
\begin{equation}\label{4}
Z^a\bar W_a\equiv\mu^\alpha v_\alpha+\bar u_{\dot\alpha}\bar\nu^{\dot\alpha}=0,\quad(Z^a\bar W_a)^\ast\equiv W^a\bar Z_a=0.
\end{equation}
The above conditions constrain the main parts of twistors $Z^a$ and $W^a$ so that they can be expressed in terms of the  coordinates $x^m$ of the points in $D=4$ real Minkowski space-time and the projectional parts
\begin{equation}
\mu^\alpha=i\bar u_{\dot\alpha}x^{\dot\alpha\alpha},\quad\nu^\alpha=i\bar v_{\dot\alpha}x^{\dot\alpha\alpha},
\end{equation}
where $x^{\dot\alpha\alpha}\equiv x^m\tilde\sigma_m^{\dot\alpha\alpha}$.
Corresponding relations for the main parts of dual twistors $\bar Z_a$ and $\bar W_a$ read
\begin{equation}
\bar\mu^{\dot\alpha}=-ix^{\dot\alpha\alpha}u_{\alpha},\quad\bar\nu^{\dot\alpha}=-ix^{\dot\alpha\alpha}v_{\alpha}.
\end{equation}
We assume that the projectional parts of the twistors form the normalized Newman-Penrose dyad
\begin{equation}\label{1}
u^\alpha v_\alpha=1,\quad\bar u^{\dot\alpha}\bar v_{\dot\alpha}=1.
\end{equation}
Such normalization conditions in the twistor notation acquire the form
\begin{equation}\label{2}
W^aI_{ab}Z^b=1,\quad\bar W_aI^{ab}\bar Z_b=1,
\end{equation}
where
\begin{equation}
I_{ab}\equiv\left(\begin{array}{cc}
0&0\\
0&\varepsilon^{\dot\alpha\dot\beta}
\end{array}\right),\quad
I^{ab}\equiv\left(\begin{array}{cc}
\varepsilon^{\alpha\beta}&0\\
0&0
\end{array}\right)
\end{equation}
are the so called asymptotic twistors\footnote{Note that asymptotic twistors are not conformally invariant.}.

The fact that twistor variables are constrained ones implies that their differentials/variations can not be arbitrary. The expressions for the admissible twistor variations compatible with the constraints (\ref{3}), (\ref{4}) and (\ref{2}) can be derived by inserting into the general twistor variation
\begin{equation}
\delta=\delta Z^a\frac{\partial}{\partial Z^a}+\delta W^a\frac{\partial}{\partial W^a}+\delta\bar Z_a\frac{\partial}{\partial\bar Z_a}+\delta\bar W_a\frac{\partial}{\partial\bar W_a}
\end{equation}
the relation
\begin{equation}\label{complete1}
\delta^a_b=I_{bc}Z^cW^a-I_{bc}W^cZ^a+\bar W_bI^{ac}\bar Z_c-\bar Z_bI^{ac}\bar W_c
\end{equation}
and performing some straightforward rearrangement of terms. The result has the form
\begin{equation}
\label{var2}
\begin{array}{rl}
\delta=&\omega(\delta)s+\bar\omega(\delta)\bar s+\tilde\omega(\delta)\tilde s+\bar{\tilde\omega}(\delta)\bar{\tilde s}+\Omega(\delta)S+\bar\Omega(\delta)\bar S\\[0.2cm]
+&\omega_{Z}(\delta)P_{Z}+\omega_{W}(\delta)P_{W}+\omega_{Z\bar W}(\delta)P_{Z\bar W}+\omega_{W\bar Z}(\delta)P_{W\bar Z}\\[0.2cm]
+&\delta(Z\bar Z)\lambda_{Z}+\delta(W\bar W)\lambda_{W}+\delta(Z\bar W)\lambda_{Z\bar W}+\delta(W\bar Z)\lambda_{W\bar Z}\\[0.2cm]
+&\frac12\delta(WIZ)g+\frac12\delta(\bar WI\bar Z)\bar g.
\end{array}
\end{equation}
Expression (\ref{var2}) contains the admissible twistor variations
\begin{equation}
\begin{array}{c}
\omega(\delta)=-\frac12(\delta ZIW+\delta WIZ)=\frac12(\delta\bar u^{\dot\alpha}\bar v_{\dot\alpha}+\delta\bar v^{\dot\alpha}\bar u_{\dot\alpha}),\\[0.2cm]
\tilde\omega(\delta)=\delta ZIZ=\bar u^{\dot\alpha}\delta\bar u_{\dot\alpha},\ \Omega(\delta)=-\delta WIW=\delta\bar v^{\dot\alpha}\bar v_{\dot\alpha},\\[0.2cm]
\omega_Z(\delta)=(Z\delta\bar Z-\delta Z\bar Z)=-2i\bar u\delta xu,\ \omega_W(\delta)=(W\delta\bar W-\delta W\bar W)=-2i\bar v\delta xv,\\[0.2cm]
\omega_{Z\bar W}(\delta)=(Z\delta\bar W-\delta Z\bar W)=-2i\bar u\delta xv,\ \omega_{W\bar Z}(\delta)=-(\omega_{Z\bar W}(\delta))^\ast.
\end{array}
\end{equation}
The summation over twistor indices and $SL(2,{\mathbb C})$ spinor indices was omitted, where it should not lead to confusion. Corresponding differential operators are given by the following expressions
\begin{equation}
\begin{array}{c}
s=Z^a\frac{\partial}{\partial Z^a}-W^a\frac{\partial}{\partial W^a},\ \tilde s=W^a\frac{\partial}{\partial Z^a},\ S=Z^a\frac{\partial}{\partial W^a},\\[0.2cm]
P_{Z}=\frac12((I\bar W)^a\frac{\partial}{\partial Z^a}-(IW)_a\frac{\partial}{\partial\bar Z_a}),\ P_{W}=\frac12((IZ)_a\frac{\partial}{\partial\bar W_a}-(I\bar Z)^a\frac{\partial}{\partial W^a}),\\[0.2cm]
P_{Z\bar W}=-\frac12((I\bar Z)^a\frac{\partial}{\partial Z^a}+(IW)_a\frac{\partial}{\partial\bar W_a}),\ P_{W\bar Z}=\frac12((I\bar W)^a\frac{\partial}{\partial W^a}+(IZ)_a\frac{\partial}{\partial\bar Z_a}),\\[0.2cm]
\lambda_{Z}=-\frac12((I\bar W)^a\frac{\partial}{\partial Z^a}+(IW)_a\frac{\partial}{\partial\bar Z_a}),\ \lambda_{W}=\frac12((I\bar Z)^a\frac{\partial}{\partial W^a}+(IZ)_a\frac{\partial}{\partial\bar W_a}),\\[0.2cm]
\lambda_{Z\bar W}=\frac12((I\bar Z)^a\frac{\partial}{\partial Z^a}-(IW)_a\frac{\partial}{\partial\bar W_a}),\ \lambda_{W\bar Z}=\frac12((IZ)_a\frac{\partial}{\partial\bar Z_a}-(I\bar W)^a\frac{\partial}{\partial W^a}),\\[0.2cm]
g=Z^a\frac{\partial}{\partial Z^a}+W^a\frac{\partial}{\partial W^a}
\end{array}
\end{equation}
and the c.c. ones. The operators $s$, $\tilde s$, $S$ satisfy the commutation relations of the $sl(2,\mathbb C)$ algebra
\begin{equation}\label{sl2c}
[s,\tilde s]=-2\tilde s,\ [s,S]=2S,\ [S,\tilde s]=s
\end{equation}
and have the following nonzero commutators with the generators $P_{Z}$, $P_{W}$, $P_{W\bar Z}$, $P_{Z\bar W}$ that correspond to translations in $D=4$ real Minkowski space-time
\begin{equation}
\begin{array}{c}
[s,P_{Z}]=-P_{Z},\ [s,P_{W}]=P_{W},\ [s,P_{W\bar Z}]=P_{W\bar Z},\ [s,P_{Z\bar W}]=-P_{Z\bar W},\\[0.2cm]
[\tilde s,P_{W}]=-P_{Z\bar W},\ [\tilde s,P_{W\bar Z}]=-P_{Z},\ [S,P_{Z}]=-P_{W\bar Z},\ [S,P_{Z\bar W}]=-P_{W}.
\end{array}
\end{equation}

We thus observe that on the constraint shell (\ref{3}), (\ref{4}), (\ref{2}) the admissible variation acquires the form
\begin{equation}
\begin{array}{rl}
\delta_{\mbox{\small admissible}}=&\omega(\delta)s+\bar\omega(\delta)\bar s+\tilde\omega(\delta)\tilde s+\bar{\tilde\omega}(\delta)\bar{\tilde s}+\Omega(\delta)S+\bar\Omega(\delta)\bar S\\[0.2cm]
+&\omega_{Z}(\delta)P_{Z}+\omega_{W}(\delta)P_{W}+\omega_{Z\bar W}(\delta)P_{Z\bar W}+\omega_{W\bar Z}(\delta)P_{W\bar Z}.
\end{array}
\end{equation}
Note also that the $sl(2,\mathbb C)$, their conjugate and translation generators forming the Poincare algebra and collectively denoted by ${\cal P}=\{s, \tilde s, S, \bar s, \bar{\tilde s}, \bar S, P_Z, P_W, P_{Z\bar W}, P_{W\bar Z}\}$, respect the constraints (\ref{3}), (\ref{4}), (\ref{2}):  ${\cal P}[\mbox{constraints (\ref{3}),(\ref{4}), (\ref{2})}]\sim\mbox{constraints (\ref{3}),(\ref{4}), (\ref{2})}$, so the theory formulated in the subspace of the twistor space defined by the constraints (\ref{3}), (\ref{4}), (\ref{2}) is invariant under the global Poincare transformations.

Among other differential operators entering (\ref{var2}) $g$ can be identified with the generator from the coset $gl(2,\mathbb C)/sl(2,\mathbb C)$ and $\lambda$'s with the translation generators in the $y$ directions of the complexified Minkowski space-time with coordinates $z^m=x^m+iy^m$. These operators do not preserve the constraints (\ref{3}), (\ref{4}), (\ref{2})
\begin{equation}
\begin{array}{c}
g(WIZ-1)=WIZ,\\[0.2cm]
\lambda_{Z}(Z^a\bar Z_a)=\lambda_{W}(W^a\bar W_a)=\lambda_{Z\bar W}(Z^a\bar W_a)=\lambda_{W\bar Z}(W^a\bar Z_a)=WIZ+\bar WI\bar Z.
\end{array}
\end{equation}
For completeness let us adduce the nonzero commutation relations of these operators with those from ${\cal P}$
\begin{equation}
\begin{array}{c}
[g,P_{Z}]=\lambda_{Z},\ [g,P_{W}]=\lambda_{W},\ [g,P_{Z\bar W}]=\lambda_{Z\bar W},\ [g,P_{W\bar Z}]=\lambda_{W\bar Z},\\[0.2cm]
[g,\lambda_{Z}]=P_{Z},\ [g,\lambda_{W}]=P_{W},\ [g,\lambda_{Z\bar W}]=P_{Z\bar W},\ [g,\lambda_{W\bar Z}]=P_{W\bar Z},\\[0.2cm]
[s,\lambda_{Z}]=-\lambda_{Z},\ [s,\lambda_{W}]=\lambda_{W},\ [s,\lambda_{Z\bar W}]=-\lambda_{Z\bar W},\ [s,\lambda_{W\bar Z}]=\lambda_{W\bar Z},\\[0.2cm]
[\tilde s,\lambda_{W}]=-\lambda_{Z\bar W},\ [\tilde s,\lambda_{W\bar Z}]=-\lambda_{Z},\ [S,\lambda_{Z}]=-\lambda_{W\bar Z},\ [S,\lambda_{Z\bar W}]=-\lambda_{W}.
\end{array}
\end{equation}
So we obtain that the admissible variation of the twistor variables is
\begin{equation}\label{dif}
\begin{array}{c}
\delta Z^a=\omega(\delta)Z^a+\tilde\omega(\delta)W^a-\frac12\omega_{Z\bar W}(\delta)I^{ab}\bar Z_b+\frac12\omega_Z(\delta)I^{ab}\bar W_b,\\[0.2cm]
\delta W^a=\Omega(\delta)Z^a-\omega(\delta)W^a-\frac12\omega_{W}(\delta)I^{ab}\bar Z_b+\frac12\omega_{W\bar Z}(\delta)I^{ab}\bar W_b,\\[0.2cm]
\delta\bar Z_a=\bar\omega(\delta)\bar Z_a+\bar{\tilde\omega}(\delta)\bar W_a+\frac12\omega_{W\bar Z}(\delta)I_{ab}Z^b-\frac12\omega_Z(\delta)I_{ab}W^b,\\[0.2cm]
\delta\bar W_a=\bar\Omega(\delta)\bar Z_a-\bar{\omega}(\delta)\bar W_a+\frac12\omega_{W}(\delta)I_{ab}Z^b-\frac12\omega_{Z\bar W}(\delta)I_{ab}W^b.
\end{array}
\end{equation}
Analogous relations hold for the differentials of twistors with the formal substitution $d\leftrightarrow\delta$.

Having defined all the necessary prerequisites we can proceed
further to define the twistor formulation of the bosonic string
action. For our purposes the most suitable starting point is the
Lorentz-harmonic formulation of the string action \cite{BZstring}
that is classically equivalent to the Polyakov one and includes
the Newman-Penrose dyad components as auxiliary fields. The action
is presented as the integral of the 2-form over the string
world sheet ${\rm M}^2$
\begin{equation}
S=\int\limits_{{\rm M}^2}L(\xi),\quad L(\xi)={\textstyle\frac{1}{2(\alpha^\prime)^{1/2}}}[e^{(+2)}\wedge(\bar udxu)-e^{(-2)}\wedge(\bar vdxv)]+{\textstyle\frac{c}{2}}e^{(-2)}\wedge e^{(+2)},
\end{equation}
where $e^{(\pm2)}=d\xi^\mu e^{(\pm2)}_\mu$ are the components of the world-sheet zweibein 1-form, $e=\frac12e^{(-2)}\wedge e^{(+2)}$ is the zweibein determinant, $\alpha^\prime$ is the constant of dimension $L^2$ (the Regge slope parameter) and $c$ is the dimensionless constant so that the string tension is $T=\frac{1}{2c\alpha^\prime}$. Here and below $\wedge$ stands for the wedge product of differential forms.

The clue to the construction of the twistor formulation of the bosonic string action is provided by the formulae
\begin{equation}\label{bridge}
\bar udxu=\frac{i}{2}\omega_Z(d)=\frac{i}{2}(Zd\bar Z-dZ\bar Z),\quad\bar vdxv=\frac{i}{2}\omega_W(d)=\frac{i}{2}(Wd\bar W-dW\bar W).
\end{equation}
These relations constitute the bridge between the space-time and twistor formulation of the bosonic string action
\begin{equation}\label{tw}
S=\int\limits_{{\rm M}^2}L(\xi),\quad L(\xi)={\textstyle\frac{i}{4(\alpha^\prime)^{1/2}}}[e^{(+2)}\wedge\omega_Z(d)-
e^{(-2)}\wedge\omega_W(d)]+{\textstyle\frac{c}{2}}e^{(-2)}\wedge e^{(+2)}.
\end{equation}

We can now proceed further to study the equations of motion
following from the twistor formulation of the string action
(\ref{tw}). The above defined relations for differentials/variations of the
twistor variables and the utilization of the known formula
\begin{equation}\label{seminal}
\delta L=i_\delta(dL)+d(i_\delta L),
\end{equation}
where $i_\delta$ can be regarded as the formal substitution of the the variation symbol for the differential,
allow to write down the general variation of the action functional (\ref{tw})
\begin{equation}\label{var1}
\begin{array}{rl}
\delta S=&\int\limits_{{\rm M}^2}{\textstyle\frac{i}{4(\alpha^\prime)^{1/2}}}e^{(+2)}\wedge[(\omega+\bar\omega)(\delta)\omega_Z(d)-(\omega+\bar\omega)(d)\omega_Z(\delta)-\tilde\omega(d)\omega_{W\bar Z}(\delta)\\[0.2cm]
+&\tilde\omega(\delta)\omega_{W\bar Z}(d)-\bar{\tilde\omega}(d)\omega_{Z\bar W}(\delta)+\bar{\tilde\omega}(\delta)\omega_{Z\bar W}(d)]+{\textstyle\frac{i}{4(\alpha^\prime)^{1/2}}}(\delta e^{(+2)}\wedge\omega_Z(d)-de^{(+2)}\omega_Z(\delta))\\[0.2cm]
-&{\textstyle\frac{i}{4(\alpha^\prime)^{1/2}}}e^{(-2)}\wedge[(\omega+\bar\omega)(d)\omega_W(\delta)-(\omega+\bar\omega)(\delta)\omega_W(d)-\Omega(d)\omega_{Z\bar W}(\delta)\\[0.2cm]
+&\Omega(\delta)\omega_{Z\bar W}(d)-\bar\Omega(d)\omega_{W\bar Z}(\delta)+\bar\Omega(\delta)\omega_{W\bar Z}(d)]+{\textstyle\frac{i}{4(\alpha^\prime)^{1/2}}}(de^{(-2)}\omega_W(\delta)-\delta e^{(-2)}\wedge\omega_W(d))\\[0.2cm]
+&\frac{c}{2}(e^{(-2)}\wedge\delta e^{(+2)}+\delta e^{(-2)}\wedge e^{(+2)}).
\end{array}
\end{equation}
When deriving the expression (\ref{var1}) we used the fact that
\begin{equation}
\begin{array}{c}
d\omega_Z=-(\omega+\bar\omega)(d)\wedge\omega_Z(d)-\tilde\omega(d)\wedge\omega_{W\bar Z}(d)-\bar{\tilde\omega}(d)\wedge\omega_{Z\bar W}(d),\\[0.2cm]
d\omega_W=(\omega+\bar\omega)(d)\wedge\omega_W(d)-\Omega(d)\wedge\omega_{Z\bar W}(d)-\bar{\Omega}(d)\wedge\omega_{W\bar Z}(d)
\end{array}
\end{equation}
and have not taken into account the contribution of the second summand of (\ref{seminal}) thus confining ourselves to the closed string case. The equations of motion following from the nullification of the variation (\ref{var1}) are given by the expressions
\begin{equation}\label{19}
\frac{\delta S}{\delta\omega}=e^{(+2)}\wedge\omega_Z(d)+e^{(-2)}\wedge\omega_W(d)=0,
\end{equation}
\begin{equation}\label{20}
\frac{\delta S}{\delta\tilde\omega}=e^{(+2)}\wedge\omega_{W\bar Z}(d)=0,
\end{equation}
\begin{equation}\label{21}
\frac{\delta S}{\delta\Omega}=e^{(-2)}\wedge\omega_{Z\bar W}(d)=0,
\end{equation}
\begin{equation}\label{22}
\frac{\delta S}{\delta\omega_Z}=de^{(+2)}+e^{(+2)}\wedge(\omega+\bar\omega)(d)=0,
\end{equation}
\begin{equation}\label{23}
\frac{\delta S}{\delta\omega_W}=de^{(-2)}-e^{(-2)}\wedge(\omega+\bar\omega)(d)=0,
\end{equation}
\begin{equation}\label{24}
\frac{\delta S}{\delta\omega_{W\bar Z}}=e^{(+2)}\wedge\tilde\omega(d)-e^{(-2)}\wedge\bar\Omega(d)=0
\end{equation}
and the c.c. ones together with the equations of motion for the zweibein 1-form
\begin{equation}\label{rheo}
e^{(-2)}={\textstyle\frac{i}{2c(\alpha^\prime)^{1/2}}}\omega_Z(d),\quad e^{(+2)}={\textstyle\frac{i}{2c(\alpha^\prime)^{1/2}}}\omega_W(d).
\end{equation}
Eqs. (\ref{20}) and (\ref{21}) and their c.c. are equivalent to the following ones
\begin{equation}\label{ort}
\omega_{Z\bar W}(d)=\omega_{W\bar Z}(d)=0.
\end{equation}
These relations (\ref{rheo}), (\ref{ort}) in view of (\ref{bridge}) are the twistor analogues of the rheotropic conditions\footnote{Such terminology originated from the group-manifold approach \cite{gma} to (super)gravity theories was adapted to super $p-$branes in \cite{BSV}.} of the Lorentz-harmonic formulation that relate the space-time coordinate differentials to the world-sheet zweibein components and can be interpreted as defining the induced zweibein\footnote{Note that the  substitution of the expressions (\ref{rheo}) back in to the action (\ref{tw}) leads to the twistor representation of the string action considered in \cite{GZ}.}. It follows immediately from (\ref{rheo}) that equation (\ref{19}) turns into identity thus exhibiting $SO(1,1)\times SO(2)$ gauge invariance of the action (\ref{tw}) (see \cite{BZstring} for details).
Eqs. (\ref{22}) and (\ref{23}) represent the zero torsion conditions for the induced zweibein and are satisfied identically provided (\ref{rheo}) and (\ref{ort}) are used. The remaining Eq.(\ref{24}) together with the only nontrivial relation following from the integrability conditions for the twistor differentials (\ref{dif})
\begin{equation}
e^{(+2)}\wedge\tilde\omega(d)+e^{(-2)}\wedge\bar\Omega(d)=0
\end{equation}
yield that
\begin{equation}\label{28}
e^{(+2)}\wedge\tilde\omega(d)=e^{(-2)}\wedge\bar\Omega(d)=0
\end{equation}
and the c.c. relations. So we observe that the only nontrivial equations of motion that follow from the string action (\ref{tw}) are (\ref{rheo}), (\ref{ort}) and (\ref{28}).

Gained experience in studying the twistor formulation of the bosonic string action can now be used when considering the supertwistor representation of the superstring action.

\section{Supertwistor formulation of superstring action}

The $SU(2,2|1)$ supertwistor variables are defined as follows
\begin{equation}
{\cal Z}^A\equiv(\mu^\alpha, \bar u_{\dot\alpha}, \bar\eta),\quad{\cal W}^A\equiv(\nu^\alpha, \bar v_{\dot\alpha}, \bar\zeta)
\end{equation}
and their duals are given by
\begin{equation}
\bar{\cal Z}_A\equiv(u_{\alpha}, \bar\mu^{\dot\alpha}, \eta),\quad\bar{\cal W}_A\equiv(v_{\alpha}, \bar\nu^{\dot\alpha}, \zeta).
\end{equation}
Supersymmetric generalizations of the zero norm and incidence conditions read
\begin{equation}\label{supernorm}
\begin{array}{c}
{\cal Z}^A\bar{\cal Z}_A\equiv\mu^\alpha u_\alpha+\bar u_{\dot\alpha}\bar\mu^{\dot\alpha}-\bar\eta\eta=0,\quad{\cal W}^A\bar{\cal W}_A\equiv\nu^\alpha v_\alpha+\bar v_{\dot\alpha}\bar\nu^{\dot\alpha}-\bar\zeta\zeta=0;\\[0.2cm]
{\cal Z}^A\bar{\cal W}_A\equiv\mu^\alpha v_\alpha+\bar u_{\dot\alpha}\bar\nu^{\dot\alpha}-\bar\eta\zeta=0,\quad{\cal W}^A\bar{\cal Z}_A\equiv\nu^\alpha u_\alpha+\bar v_{\dot\alpha}\bar\mu^{\dot\alpha}-\bar\zeta\eta=0.
\end{array}
\end{equation}
Their imposition allows to find the expressions for the main $\mu^\alpha$, $\nu^\alpha$ and Grassmann-odd $\eta$, $\zeta$ components of the supertwistors in terms of the $N=1$ superspace coordinates $x_{\alpha\dot\alpha}$, $\theta^\alpha$, $\bar\theta^{\dot\alpha}$ and the Newman-Penrose dyad components (\ref{1})
\begin{equation}
\begin{array}{c}
\mu^\alpha=i\bar u_{\dot\alpha}x^{\dot\alpha\alpha}+\theta^\alpha\bar\eta,\
\bar\mu^{\dot\alpha}=-ix^{\dot\alpha\alpha}u_{\alpha}+\eta\bar\theta^{\dot\alpha},\ \bar\eta=2\bar u^{\dot\alpha}\bar\theta_{\dot\alpha},\ \eta=2u^\alpha\theta_\alpha;\\[0.2cm]
\nu^\alpha=i\bar v_{\dot\alpha}x^{\dot\alpha\alpha}+\theta^\alpha\bar\zeta,\
\bar\nu^{\dot\alpha}=-ix^{\dot\alpha\alpha}v_{\alpha}+\zeta\bar\theta^{\dot\alpha},\ \bar\zeta=2\bar v^{\dot\alpha}\bar\theta_{\dot\alpha},\ \zeta=2v^\alpha\theta_\alpha.
\end{array}
\end{equation}

To find the supertwistor variation consistent with (\ref{supernorm}) consider an arbitrary variation in the supertwistor space
\begin{equation}\label{susyvar}
\delta=\delta{\cal Z}^A\frac{\partial}{\partial{\cal Z}^A}+\delta{\cal W}^A\frac{\partial}{\partial{\cal W}^A}+\delta\bar{\cal Z}_A\frac{\partial}{\partial\bar{\cal Z}_A}+\delta\bar{\cal W}_A\frac{\partial}{\partial\bar{\cal W}_A}.
\end{equation}
To single out the part of this general variation that respects the constraints (\ref{supernorm}) introduce the unit matrix in (\ref{susyvar})
\begin{equation}\label{susyunit}
\begin{array}{rl}
\delta^A_B=&{\cal W}^A({\cal IZ})_B-{\cal Z}^A({\cal IW})_B+({\cal I}\bar{\cal Z})^A\bar{\cal W}_B-({\cal I}\bar{\cal W})^A\bar{\cal Z}_B\\[0.2cm]
+&({\cal I}\bar{\cal Z})^A({\cal IZ})_B\zeta\bar\zeta+({\cal I}\bar{\cal W})^A({\cal IW})_B\eta\bar\eta-({\cal I}\bar{\cal Z})^A({\cal IW})_B\zeta\bar\eta-({\cal I}\bar{\cal W})^A({\cal IZ})_B\eta\bar\zeta\\[0.2cm]
+&({\cal JZ})^A({\cal IW})_B-({\cal JW})^A({\cal IZ})_B+({\cal I}\bar{\cal W})^A({\cal J}\bar{\cal Z})_B-({\cal I}\bar{\cal Z})^A({\cal J}\bar{\cal W})_B+{\cal J}^A_B,
\end{array}
\end{equation}
where
\begin{equation}
{\cal I}_{AB}\equiv\left(\begin{array}{ccc}
0&0&0\\
0&\varepsilon^{\dot\alpha\dot\beta}&0\\
0&0&0
\end{array}\right),\quad
{\cal I}^{AB}\equiv\left(\begin{array}{ccc}
\varepsilon^{\alpha\beta}&0&0\\
0&0&0\\
0&0&0
\end{array}\right)
\end{equation}
are the supersymmetric extensions of the asymptotic twistors. The matrix
\begin{equation}
{\cal J}^{A}_{B}
\equiv\left(\begin{array}{ccc}
0&0&0\\
0&0&0\\
0&0&1
\end{array}\right)
\end{equation}
is used to single out the Grassmann-odd component of the supertwistor.
The expression (\ref{susyunit}) is the supersymmetric counterpart of the relation (\ref{complete1}). Then the variation (\ref{susyvar}) can be brought to the form
\begin{equation}\label{susyvar2}
\begin{array}{rl}
\delta=&\omega(\delta)s+\bar\omega(\delta)\bar s+\tilde\omega(\delta)\tilde s+\bar{\tilde\omega}(\delta)\bar{\tilde s}+\Omega(\delta)S+\bar\Omega(\delta)\bar S\\[0.2cm]
+&\omega_{\cal Z}(\delta)P_{\cal Z}+\omega_{\cal W}(\delta)P_{\cal W}+\omega_{{\cal Z}\bar{\cal W}}(\delta)P_{{\cal Z}\bar{\cal W}}+\omega_{{\cal W}\bar{\cal Z}}(\delta)P_{{\cal W}\bar{\cal Z}}\\[0.2cm]
+&{\cal D}(\delta)\bar\eta Q_{\bar\eta}+{\cal D}(\delta)\eta Q_{\eta}+{\cal D}(\delta)\bar\zeta Q_{\bar\zeta}+{\cal D}(\delta)\zeta Q_\zeta\\[0.2cm]
+&\delta({\cal Z}\bar{\cal Z})\lambda_{\cal Z}+\delta({\cal W}\bar{\cal W})\lambda_{\cal W}+\delta({\cal Z}\bar{\cal W})\lambda_{{\cal Z}\bar{\cal W}}+\delta({\cal W}\bar{\cal Z})\lambda_{{\cal W}\bar{\cal Z}}\\[0.2cm]
+&\frac12\delta({\cal WIZ})g+\frac12\delta(\bar{\cal W}{\cal I}\bar{\cal Z})\bar g,
\end{array}
\end{equation}
where
$\omega_{\cal Z}(\delta)={\cal Z}\delta\bar{\cal Z}-\delta{\cal Z}\bar{\cal Z}$, $\omega_{\cal W}(\delta)={\cal W}\delta\bar{\cal W}-\delta{\cal W}\bar{\cal W}$, $\omega_{{\cal Z}\bar{\cal W}}(\delta)={\cal Z}\delta\bar{\cal W}-\delta{\cal Z}\bar{\cal W}$, $\omega_{{\cal W}\bar{\cal Z}}=-(\omega_{{\cal Z}\bar{\cal W}})^\ast$, ${\cal D}(\delta)\bar\eta=2\bar u^{\dot\alpha}\delta\bar\theta_{\dot\alpha}=\delta\bar\eta-\omega(\delta)\bar\eta-\tilde\omega(\delta)\bar\zeta$, ${\cal D}(\delta)\bar\zeta=2\bar v^{\dot\alpha}\delta\bar\theta_{\dot\alpha}=\delta\bar\zeta+\omega(\delta)\bar\zeta-\Omega(\delta)\bar\eta$
are the admissible variations respecting the constraints (\ref{supernorm}). Corresponding differential forms are the Cartan forms invariant under the $SL(2,\mathbb C)$ as well as translations and supersymmetry transformations. Differential operators that enter (\ref{susyvar2}) are given by the following expressions
\begin{equation}
\begin{array}{c}
s=\mathcal Z^A\frac{\partial}{\partial\mathcal Z^A}-\mathcal W^A\frac{\partial}{\partial\mathcal W^A},\ \tilde s=\mathcal W^A\frac{\partial}{\partial\mathcal Z^A},\ S=\mathcal Z^A\frac{\partial}{\partial\mathcal W^A},\\[0.2cm]
P_{\mathcal Z}=\frac12((\mathcal I\bar{\mathcal W})^A\frac{\partial}{\partial\mathcal Z^A}-(\mathcal{IW})_A\frac{\partial}{\partial\bar{\mathcal Z}_A}),\ P_{\mathcal W}=\frac12((\mathcal{IZ})_A\frac{\partial}{\partial\bar{\mathcal W}_A}-(\mathcal I\bar{\mathcal Z})^A\frac{\partial}{\partial\mathcal W^A}),\\[0.2cm]
P_{{\cal Z}\bar{\cal W}}=-\frac12((\mathcal I\bar{\cal Z})^A\frac{\partial}{\partial\mathcal Z^A}+(\mathcal{IW})_A\frac{\partial}{\partial\bar{\cal W}_A}),\ P_{{\cal W}\bar{\cal Z}}=\frac12((\mathcal I\bar{\cal W})^A\frac{\partial}{\partial\mathcal W^A}+(\mathcal{IZ})_A\frac{\partial}{\partial\bar{\cal Z}_A}),\\[0.2cm]
Q_{\eta}=\frac{\partial}{\partial\eta}+\bar\eta(\mathcal{IW})_A\frac{\partial}{\partial\bar{\cal Z}_A}-\bar\zeta(\mathcal{IZ})_A\frac{\partial}{\partial\bar{\cal Z}_A},\ Q_{\bar\eta}=\frac{\partial}{\partial\bar\eta}-\eta(\mathcal I\bar{\cal W})^A\frac{\partial}{\partial\mathcal Z^A}+\zeta(\mathcal I\bar{\cal Z})^A\frac{\partial}{\partial\mathcal Z^A},\\[0.2cm]
Q_{\zeta}=\frac{\partial}{\partial\zeta}+\bar\eta(\mathcal{IW})_A\frac{\partial}{\partial\bar{\cal W}_A}-\bar\zeta(\mathcal{IZ})_A\frac{\partial}{\partial\bar{\cal W}_A},\ Q_{\bar\zeta}=\frac{\partial}{\partial\bar\zeta}-\eta(\mathcal I\bar{\cal W})^A\frac{\partial}{\partial\mathcal W^A}+\zeta(\mathcal I\bar{\cal Z})^A\frac{\partial}{\partial\mathcal W^A},\\[0.2cm]
\lambda_{\cal Z}=-\frac12((\mathcal I\bar{\cal W})^A\frac{\partial}{\partial\mathcal Z^A}+(\mathcal{IW})_A\frac{\partial}{\partial\bar{\cal Z}_A}),\ \lambda_{\cal W}=\frac12((\mathcal I\bar{\cal Z})^A\frac{\partial}{\partial\mathcal W^A}+(\mathcal{IZ})_A\frac{\partial}{\partial\bar{\cal W}_A}),\\[0.2cm]
\lambda_{\mathcal Z\bar{\cal W}}=\frac12((\mathcal I\bar{\cal Z})^A\frac{\partial}{\partial\mathcal Z^A}-(\mathcal{IW})_A\frac{\partial}{\partial\bar{\cal W}_A}),\ \lambda_{\mathcal W\bar{\cal Z}}=\frac12((\mathcal{IZ})_A\frac{\partial}{\partial\bar{\cal Z}_A}-(\mathcal I\bar{\cal W})^A\frac{\partial}{\partial\cal W^A}),\\[0.2cm]
g\equiv\mathcal Z^A\frac{\partial}{\partial\mathcal Z^A}+\mathcal W^A\frac{\partial}{\partial\mathcal W^A}-\bar\eta Q_{\bar\eta}-\bar\zeta Q_{\bar\zeta}
\end{array}
\end{equation}
and the corresponding c.c. ones. The operators $s$, $\tilde s$, $S$ satisfy the commutation relations of the $sl(2,\mathbb C)$ algebra
(\ref{sl2c})
and have the following nonzero commutators with the supersymmetry $Q_\eta$, $Q_{\bar\eta}$, $Q_\zeta$, $Q_{\bar\zeta}$ and translation $P_{\mathcal Z}$, $P_{\mathcal W}$, $P_{\mathcal W\bar{\cal Z}}$, $P_{\mathcal Z\bar{\cal W}}$ generators
\begin{equation}
\begin{array}{c}
[s,Q_{\bar\eta}]=-Q_{\bar\eta},\ [s,Q_{\bar\zeta}]=Q_{\bar\zeta},\ [S,Q_{\bar\eta}]=-Q_{\bar\zeta},\ [\tilde s,Q_{\bar\zeta}]=-Q_{\bar\eta},\\[0.2cm]
[s,P_{\mathcal Z}]=-P_{\mathcal Z},\ [s,P_{\mathcal W}]=P_{\mathcal W},\ [s,P_{\mathcal W\bar{\cal Z}}]=P_{\mathcal W\bar{\cal Z}},\ [s,P_{\mathcal Z\bar{\cal W}}]=-P_{\mathcal Z\bar{\cal W}},\\[0.2cm]
[\tilde s,P_{\mathcal W}]=-P_{\mathcal Z\bar{\cal W}},\ [\tilde s,P_{\mathcal W\bar{\cal Z}}]=-P_{\mathcal Z},\ [S,P_{\mathcal Z}]=-P_{\mathcal W\bar{\cal Z}},\ [S,P_{\mathcal Z\bar{\cal W}}]=-P_{\mathcal W}.
\end{array}
\end{equation}
In turn $Q$ and $P$ operators provide the supertwistor realization of the $N=1$ supersymmetry algebra
\begin{equation}
\{Q_\eta,Q_{\bar\eta}\}=-2P_{\mathcal Z},\ \{Q_\zeta,Q_{\bar\zeta}\}=-2P_{\mathcal W},\ \{Q_\eta,Q_{\bar\zeta}\}=-2P_{\mathcal W\bar{\cal Z}},\ \{Q_{\zeta},Q_{\bar\eta}\}=-2P_{\mathcal Z\bar{\cal W}}.
\end{equation}

Hence, when the constraints (\ref{supernorm}) are imposed, the variation (\ref{susyvar2}) of the supertwistors reduces to
\begin{equation}
\begin{array}{rl}
\delta_{\mbox{\small admissible}}=&\omega(\delta)s+\bar\omega(\delta)\bar s+\tilde\omega(\delta)\tilde s+\bar{\tilde\omega}(\delta)\bar{\tilde s}+\Omega(\delta)S+\bar\Omega(\delta)\bar S\\[0.2cm]
+&\omega_{\cal Z}(\delta)P_{\cal Z}+\omega_{\cal W}(\delta)P_{\cal W}+\omega_{{\cal Z}\bar{\cal W}}(\delta)P_{{\cal Z}\bar{\cal W}}+\omega_{{\cal W}\bar{\cal Z}}(\delta)P_{{\cal W}\bar{\cal Z}}\\[0.2cm]
+&{\cal D}(\delta)\bar\eta Q_{\bar\eta}+{\cal D}(\delta)\eta Q_{\eta}+{\cal D}(\delta)\bar\zeta Q_{\bar\zeta}+{\cal D}(\delta)\zeta Q_\zeta
\end{array}
\end{equation}
that is the most general variation consistent with the constraints (\ref{supernorm}).

On the other hand, the $N=1$ superPoincare generators preserve the supertwistor constraints
\begin{equation}
\begin{array}{c}
S{\cal P}=\{s, \tilde s, S, \bar s, \bar{\tilde s}, \bar S, P_{\mathcal Z}, P_{\mathcal W}, P_{\mathcal Z\bar{\cal W}}, P_{\mathcal W\bar{\cal Z}}, Q_\eta, Q_{\bar\eta}, Q_\zeta, Q_{\bar\zeta}\}:\\[0.2cm]
S{\cal P}[\mbox{constraints (\ref{supernorm})}]\sim\mbox{constraints (\ref{supernorm})}.
\end{array}
\end{equation}
So the global symmetry of the model formulated in the subspace of the supertwistor space defined by the constraints (\ref{supernorm}) consists of $N=1$ superPoincare transformations.

Among remaining differential operators entering (\ref{susyvar2}) $g\in gl(2,\mathbb C)/sl(2,\mathbb C)$, while $\lambda$'s generate the translations in the $y$ directions of the complexified superspace parametrized by the bosonic coordinates $z^m=x^m+iy^m$. These operators do not preserve the constraints (\ref{supernorm}) as is illustrated by the relations
\begin{equation}
\begin{array}{c}
g(\mathcal{WIZ}-1)=\mathcal{WIZ},\\[0.2cm]
\lambda_{\mathcal Z}(\mathcal Z^A\bar{\cal Z}_A)=\lambda_{\mathcal W}(\mathcal W^A\bar{\cal W}_A)=\lambda_{\mathcal Z\bar{\cal W}}(\mathcal Z^A\bar{\cal W}_A)=\lambda_{\mathcal W\bar{\cal Z}}(\mathcal W^A\bar{\cal Z}_A)=\mathcal{WIZ}+\bar{\cal W}\bar{\cal I}\bar{\cal Z}
\end{array}
\end{equation}
and obey the following nonzero commutation relations with the $S{\cal P}$ generators
\begin{equation}
\begin{array}{c}
[Q_\eta,g]=\bar\eta\lambda_{\mathcal Z}+\bar\zeta\lambda_{\mathcal W\bar{\cal Z}},\ [Q_\zeta,g]=\bar\eta\lambda_{\mathcal Z\bar{\cal W}}+\bar\zeta\lambda_{\mathcal W},\\[0.2cm]
[g,P_{\mathcal Z}]=\lambda_{\mathcal Z},\ [g,P_{\mathcal W}]=\lambda_{\mathcal W},\ [g,P_{\mathcal Z\bar{\cal W}}]=\lambda_{\mathcal Z\bar{\cal W}},\ [g,P_{\mathcal W\bar{\cal Z}}]=\lambda_{\mathcal W\bar{\cal Z}},\\[0.2cm]
[g,\lambda_{\mathcal Z}]=P_{\mathcal Z},\ [g,\lambda_{\mathcal W}]=P_{\mathcal W},\ [g,\lambda_{\mathcal Z\bar{\cal W}}]=P_{\mathcal Z\bar{\cal W}},\ [g,\lambda_{\mathcal W\bar{\cal Z}}]=P_{\mathcal W\bar{\cal Z}},\\[0.2cm]
[s,\lambda_{\mathcal Z}]=-\lambda_{\mathcal Z},\ [s,\lambda_{\mathcal W}]=\lambda_{\mathcal W},\ [s,\lambda_{\mathcal Z\bar{\cal W}}]=-\lambda_{\mathcal Z\bar{\cal W}},\ [s,\lambda_{\mathcal W\bar{\cal Z}}]=\lambda_{\mathcal W\bar{\cal Z}},\\[0.2cm]
[\tilde s,\lambda_{\mathcal W}]=-\lambda_{\mathcal Z\bar{\cal W}},\ [\tilde s,\lambda_{\mathcal W\bar{\cal Z}}]=-\lambda_{\mathcal Z},\ [S,\lambda_{\mathcal Z}]=-\lambda_{\mathcal W\bar{\cal Z}},\ [S,\lambda_{\mathcal Z\bar{\cal W}}]=-\lambda_{\mathcal W}.
\end{array}
\end{equation}

The above analysis reveals that the admissible variation of the supertwistors that generalizes nonsupersymmetric expression (\ref{dif}) is
\begin{equation}\label{33}
\begin{array}{rl}
\delta{\cal Z}^A=&\omega(\delta){\cal Z}^A+\tilde\omega(\delta){\cal W}^A-(\frac12\omega_{{\cal Z}\bar{\cal W}}(\delta)+\zeta{\cal D}(\delta)\bar\eta){\cal I}^{AB}\bar{\cal Z}_B\\[0.2cm]
+&(\frac12\omega_{\cal Z}(\delta)+\eta{\cal D}(\delta)\bar\eta){\cal I}^{AB}\bar{\cal W}_B+{\cal J}^{A}_{B}\delta\mathcal Z^B,\\[0.2cm]
\delta{\cal W}^A=&\Omega(\delta){\cal Z}^A-\omega(\delta){\cal W}^A-(\frac12\omega_{\cal W}(\delta)+\zeta{\cal D}(\delta)\bar\zeta){\cal I}^{AB}\bar{\cal Z}_B\\[0.2cm]
+&(\frac12\omega_{{\cal W}\bar{\cal Z}}(\delta)+\eta{\cal D}(\delta)\bar\zeta){\cal I}^{AB}\bar{\cal W}_B+{\cal J}^{A}_{B}\delta\mathcal W^B.
\end{array}
\end{equation}
The same relations and their c.c. hold for supertwistor differentials with the substitution $d\leftrightarrow\delta$.

The $N=1$ superstring action \cite{GSW} is represented as the sum of the kinetic and the Wess-Zumino terms
\begin{equation}\label{sust}
S^{N=1}=S^{N=1}_{kin}+S^{N=1}_{WZ},
\end{equation}
where the kinetic term in the Lorentz-harmonic formulation \cite{BZstring} is
\begin{equation}
S^{N=1}_{kin}=\int\limits_{{\rm M}^2}\left({\textstyle\frac{1}{2(\alpha^\prime)^{1/2}}}[e^{(+2)}\wedge(\bar u\omega(d)u)-e^{(-2)}\wedge(\bar v\omega(d)v)]+{\textstyle\frac{c}{2}}e^{(-2)}\wedge e^{(+2)}\right)
\end{equation}
where $\omega_{\alpha\dot\alpha}(d)=dx_{\alpha\dot\alpha}+2id\theta_\alpha\bar\theta_{\dot\alpha}-2i\theta_\alpha d\bar\theta_{\dot\alpha}$ is the supersymmetric Cartan 1-form, 
and the Wess-Zumino term has the form
\begin{equation}
S^{N=1}_{WZ}=\frac{is}{c\alpha^\prime}\int\limits_{{\rm M}^2}\omega^{\dot\alpha\alpha}(d)\wedge(d\theta_\alpha\bar\theta_{\dot\alpha}-\theta_\alpha d\bar\theta_{\dot\alpha}).
\end{equation}
Numerical factor $s$ can take values $\pm1$ in consistency with the $\kappa$-invariance of the action (\ref{sust}).

The supertwistor representation for the kinetic term is obtained with the aid of relations
\begin{equation}\label{38}
\bar u\omega(d)u=\frac{i}{2}\omega_{\cal Z}(d)=\frac{i}{2}({\cal Z}d\bar{\cal Z}-d{\cal Z}\bar{\cal Z}),\quad
\bar v\omega(d)v=\frac{i}{2}\omega_{\cal W}(d)=\frac{i}{2}({\cal W}d\bar{\cal W}-d{\cal W}\bar{\cal W})
\end{equation}
that generalize (\ref{bridge}) and has the form
\begin{equation}\label{39}
S^{N=1}_{kin}=\int\limits_{{\rm M}^2}\left({\textstyle\frac{i}{4(\alpha^\prime)^{1/2}}}[e^{(+2)}\wedge\omega_{\cal Z}(d)-e^{(-2)}\wedge\omega_{\cal W}(d)]+{\textstyle\frac{c}{2}}e^{(-2)}\wedge e^{(+2)}\right).
\end{equation}
Via the completeness conditions for the Newman-Penrose dyad
\begin{equation}
u^\alpha v_\beta-v^\alpha u_\beta=\delta^\alpha_\beta,\quad\bar u^{\dot\alpha}\bar v_{\dot\beta}-\bar v^{\dot\alpha}\bar u_{\dot\beta}=\delta^{\dot\alpha}_{\dot\beta}
\end{equation}
its components can be introduced into the Wess-Zumino term that acquires the form
\begin{equation}\label{41}
S^{N=1}_{WZ}=-\frac{s}{8c\alpha^\prime}\int\limits_{{\rm M}^2}\left(\omega_{\cal Z}(d)\wedge\omega_\zeta(d)+\omega_{\cal W}(d)\wedge\omega_\eta(d)-\omega_{{\cal W}\bar{\cal Z}}(d)\wedge\omega_{\zeta\bar\eta}(d)-\omega_{{\cal Z}\bar{\cal W}}(d)\wedge\omega_{\eta\bar\zeta}(d)\right),
\end{equation}
where
\begin{equation}
\begin{array}{c}
\omega_\eta(d)={\cal D}\eta\bar\eta-\eta{\cal D}\bar\eta,\quad\omega_\zeta(d)={\cal D}\zeta\bar\zeta-\zeta{\cal D}\bar\zeta,\\[0.2cm]
\omega_{\zeta\bar\eta}(d)={\cal D}\zeta\bar\eta-\zeta{\cal D}\bar\eta,\quad\omega_{\eta\bar\zeta}(d)=-(\omega_{\zeta\bar\eta})^\ast
\end{array}
\end{equation}
and the covariant differentials of Grassmann-odd supertwistor components are defined as
\begin{equation}
\begin{array}{c}
{\cal D}\eta=d\eta-\bar\omega(d)\eta-\bar{\tilde\omega}(d)\zeta,\quad {\cal D}\bar\eta=d\bar\eta-\omega(d)\bar\eta-{\tilde\omega}(d)\bar\zeta;\\[0.2cm]
{\cal D}\zeta=d\zeta+\bar\omega(d)\zeta-\bar\Omega(d)\eta,\quad {\cal D}\bar\zeta=d\bar\zeta+\omega(d)\bar\zeta-\Omega(d)\bar\eta.
\end{array}
\end{equation}

To derive the $N=1$ superstring equations of motion in the
supertwistor formulation it is convenient to utilize the
expressions for the differentials of the supersymmetric 1-forms
\begin{equation}
\begin{array}{c}
d\omega_{\cal Z}=-(\omega+\bar\omega)(d)\wedge\omega_{\cal Z}(d)-\tilde\omega(d)\wedge\omega_{{\cal W}\bar{\cal Z}}(d)-\bar{\tilde\omega}(d)\wedge\omega_{{\cal Z}\bar{\cal W}}(d)+2{\cal D}\bar\eta\wedge{\cal D}\eta,\\[0.2cm]
d\omega_{\cal W}=(\omega+\bar\omega)(d)\wedge\omega_{\cal W}(d)-\Omega(d)\wedge\omega_{{\cal Z}\bar{\cal W}}(d)-\bar{\Omega}(d)\wedge\omega_{{\cal W}\bar{\cal Z}}(d)+2{\cal D}\bar\zeta\wedge{\cal D}\zeta,\\[0.2cm]
d\omega_{{\cal W}\bar{\cal Z}}=(\omega-\bar\omega)(d)\wedge\omega_{{\cal W}\bar{\cal Z}}(d)-\Omega(d)\wedge\omega_{\cal Z}(d)-\bar{\tilde\omega}(d)\wedge\omega_{\cal W}(d)+2{\cal D}\bar\zeta\wedge{\cal D}\eta,\\[0.2cm]
d\omega_{{\cal Z}\bar{\cal W}}=(\bar\omega-\omega)(d)\wedge\omega_{{\cal Z}\bar{\cal W}}(d)-\bar\Omega(d)\wedge\omega_{\cal Z}(d)-\tilde\omega(d)\wedge\omega_{\cal W}(d)+2{\cal D}\bar\eta\wedge{\cal D}\zeta,\\[0.2cm]
\end{array}
\end{equation}
as well as for the differentials of the 1-forms composed of Grassmann-odd supertwistor components
\begin{equation}
\begin{array}{c}
d\omega_\zeta=(\omega+\bar\omega)(d)\wedge\omega_\zeta(d)-\Omega(d)\wedge\omega_{\zeta\bar\eta}(d)-\bar\Omega(d)\wedge\omega_{\eta\bar\zeta}(d)+2{\cal D}\bar\zeta\wedge{\cal D}\zeta,\\[0.2cm]
d\omega_\eta=-(\omega+\bar\omega)(d)\wedge\omega_\eta(d)-\tilde\omega(d)\wedge\omega_{\eta\bar\zeta}(d)-\bar{\tilde\omega}(d)\wedge\omega_{\zeta\bar\eta}(d)+2{\cal D}\bar\eta\wedge{\cal D}\eta,\\[0.2cm]
d\omega_{\eta\bar\zeta}=(\omega-\bar\omega)(d)\wedge\omega_{\eta\bar\zeta}(d)-\Omega(d)\wedge\omega_\eta(d)-\bar{\tilde\omega}(d)\wedge\omega_\zeta(d)+2{\cal D}\bar\zeta\wedge{\cal D}\eta,\\[0.2cm]
d\omega_{\zeta\bar\eta}=(\bar\omega-\omega)(d)\wedge\omega_{\zeta\bar\eta}(d)-\bar\Omega(d)\wedge\omega_\eta(d)-\tilde\omega(d)\wedge\omega_\zeta(d)+2{\cal D}\bar\eta\wedge{\cal D}\zeta.
\end{array}
\end{equation}
Using the above expressions the exterior differential of the Wess-Zumino term in the supertwistor formulation is given by
\begin{equation}
dS^{N=1}_{WZ}=-\frac{s}{4c\alpha^\prime}\int\limits_{{\rm M}^2}(\omega_{\cal Z}(d)\wedge{\cal D}\zeta\wedge{\cal D}\bar\zeta+\omega_{\cal W}(d)\wedge{\cal D}\eta\wedge{\cal D}\bar\eta-\omega_{{\cal W}\bar{\cal Z}}(d)\wedge{\cal D}\zeta\wedge{\cal D}\bar\eta-\omega_{{\cal Z}\bar{\cal W}}(d)\wedge{\cal D}\eta\wedge{\cal D}\bar\zeta)
\end{equation}
and the action variation acquires the form
\begin{equation}
\delta S^{N=1}=\delta S^{N=1}_{kin}+\delta S^{N=1}_{WZ}
\end{equation}
with
\begin{equation}
\begin{array}{rl}
\delta S^{N=1}_{kin}=&\int\limits_{{\rm M}^2}{\textstyle\frac{i}{4(\alpha^\prime)^{1/2}}}e^{(+2)}\wedge[(\omega+\bar\omega)(\delta)\omega_{\cal Z}(d)-(\omega+\bar\omega)(d)\omega_{\cal Z}(\delta)-\tilde\omega(d)\omega_{{\cal W}\bar{\cal Z}}(\delta)\\[0.2cm]
+&\tilde\omega(\delta)\omega_{{\cal W}\bar{\cal Z}}(d)-\bar{\tilde\omega}(d)\omega_{{\cal Z}\bar{\cal W}}(\delta)+\bar{\tilde\omega}(\delta)\omega_{{\cal Z}\bar{\cal W}}(d)+2{\cal D}\bar\eta{\cal D}(\delta)\eta-2{\cal D}(\delta)\bar\eta{\cal D}\eta]\\[0.2cm]
+&{\textstyle\frac{i}{4(\alpha^\prime)^{1/2}}}(\delta e^{(+2)}\wedge\omega_{\cal Z}(d)-de^{(+2)}\omega_{\cal Z}(\delta))\\[0.2cm]
-&{\textstyle\frac{i}{4(\alpha^\prime)^{1/2}}}e^{(-2)}\wedge[(\omega+\bar\omega)(d)\omega_{\cal W}(\delta)-(\omega+\bar\omega)(\delta)\omega_{\cal W}(d)-\Omega(d)\omega_{{\cal Z}\bar{\cal W}}(\delta)\\[0.2cm]
+&\Omega(\delta)\omega_{{\cal Z}\bar{\cal W}}(d)-\bar\Omega(d)\omega_{{\cal W}\bar{\cal Z}}(\delta)+\bar\Omega(\delta)\omega_{{\cal W}\bar{\cal Z}}(d)+2{\cal D}\bar\zeta{\cal D}(\delta)\zeta-2{\cal D}(\delta)\bar\zeta{\cal D}\zeta]\\[0.2cm]
+&{\textstyle\frac{i}{4(\alpha^\prime)^{1/2}}}(de^{(-2)}\omega_{\cal W}(\delta)-\delta e^{(-2)}\wedge\omega_{\cal W}(d))\\[0.2cm]
+&\frac{c}{2}(e^{(-2)}\wedge\delta e^{(+2)}+\delta e^{(-2)}\wedge e^{(+2)})
\end{array}
\end{equation}
and
\begin{equation}
\begin{array}{rl}
\delta S^{N=1}_{WZ}=&-\frac{s}{4c\alpha^\prime}\int\limits_{{\rm M}^2}[\omega_{\cal Z}(d)\wedge{\cal D}\zeta{\cal D}(\delta)\bar\zeta-\omega_{\cal Z}(d){\cal D}(\delta)\zeta\wedge{\cal D}\bar\zeta+\omega_{\cal Z}(\delta){\cal D}\zeta\wedge{\cal D}\bar\zeta\\[0.2cm]
+&\omega_{\cal W}(d)\wedge{\cal D}\eta{\cal D}(\delta)\bar\eta-\omega_{\cal W}(d){\cal D}(\delta)\eta\wedge{\cal D}\bar\eta+\omega_{\cal W}(\delta){\cal D}\eta\wedge{\cal D}\bar\eta\\[0.2cm]
-&\omega_{{\cal W}\bar{\cal Z}}(d)\wedge{\cal D}\zeta{\cal D}(\delta)\bar\eta+\omega_{{\cal W}\bar{\cal Z}}(d){\cal D}(\delta)\zeta\wedge{\cal D}\bar\eta-\omega_{{\cal W}\bar{\cal Z}}(\delta){\cal D}\zeta\wedge{\cal D}\bar\eta\\[0.2cm]
-&\omega_{{\cal Z}\bar{\cal W}}(d)\wedge{\cal D}\eta{\cal D}(\delta)\bar\zeta+\omega_{{\cal Z}\bar{\cal W}}(d){\cal D}(\delta)\eta\wedge{\cal D}\bar\zeta-\omega_{{\cal Z}\bar{\cal W}}(\delta){\cal D}\eta\wedge{\cal D}\bar\zeta].
\end{array}
\end{equation}
Considering variations $\delta e^{(\pm2)}$, $\omega(\delta)$, $\tilde\omega(\delta)$, $\Omega(\delta)$, $\omega_{\cal Z}(\delta)$, $\omega_{\cal W}(\delta)$, $\omega_{{\cal Z}\bar{\cal W}}(\delta)$, ${\cal D}(\delta)\bar\eta$, ${\cal D}(\delta)\bar\zeta$ and c.c. ones as independent one finds the  equations of motion for the superstring in the supertwistor formulation. Namely, for the zweibein components we get
\begin{equation}
\frac{\delta S^{N=1}}{\delta e^{(+2)}}=-{\textstyle\frac{i}{4(\alpha^\prime)^{1/2}}}\omega_{\cal Z}(d)+{\textstyle\frac{c}{2}}e^{(-2)}=0,\quad\frac{\delta S^{N=1}}{\delta e^{(-2)}}={\textstyle\frac{i}{4(\alpha^\prime)^{1/2}}}\omega_{\cal W}(d)-{\textstyle\frac{c}{2}}e^{(+2)}=0
\end{equation}
leading to the supertwistor counterparts of the rheotropic conditions of the Lorentz-harmonic formulation
\begin{equation}\label{srheo}
\omega_{\cal Z}(d)=-2ic(\alpha^\prime)^{1/2}e^{(-2)},\quad\omega_{\cal W}(d)=-2ic(\alpha^\prime)^{1/2}e^{(+2)}.
\end{equation}
The equation
\begin{equation}
\frac{\delta S^{N=1}}{\delta\omega}=e^{(+2)}\wedge\omega_{\cal Z}(d)+e^{(-2)}\wedge\omega_{\cal W}(d)=0
\end{equation}
is satisfied identically provided (\ref{srheo}) is used signalizing the $SO(1,1)\times SO(2)$ gauge symmetry presence analogously to the bosonic case. The equations
\begin{equation}
\frac{\delta S^{N=1}}{\delta\tilde\omega}=e^{(+2)}\wedge\omega_{{\cal W}\bar{\cal Z}}(d)=0,\quad\frac{\delta S^{N=1}}{\delta\Omega}=e^{(-2)}\wedge\omega_{{\cal Z}\bar{\cal W}}(d)=0
\end{equation}
and c.c. ones are equivalent to the second pair of the rheotropic relations
\begin{equation}\label{srheo2}
\omega_{{\cal Z}\bar{\cal W}}(d)=\omega_{{\cal W}\bar{\cal Z}}(d)=0.
\end{equation}
Other bosonic equations read
\begin{equation}\label{55}
\begin{array}{c}
\frac{\delta S^{N=1}}{\delta\omega_{\cal Z}}=[de^{(+2)}+e^{(+2)}\wedge(\omega+\bar\omega)(d)]-\frac{is}{c(\alpha^\prime)^{1/2}}{\cal D}\zeta\wedge{\cal D}\bar\zeta=0,\\[0.2cm]
\frac{\delta S^{N=1}}{\delta\omega_{\cal W}}=[de^{(-2)}-e^{(-2)}\wedge(\omega+\bar\omega)(d)]+\frac{is}{c(\alpha^\prime)^{1/2}}{\cal D}\eta\wedge{\cal D}\bar\eta=0,\\[0.2cm]
\frac{\delta S^{N=1}}{\delta\omega_{{\cal W}\bar{\cal Z}}}=e^{(-2)}\wedge\bar\Omega(d)-e^{(+2)}\wedge\tilde\omega(d)-\frac{is}{c(\alpha^\prime)^{1/2}}{\cal D}\zeta\wedge{\cal D}\bar\eta=0.
\end{array}
\end{equation}
The last equation should be complemented by its c.c. Note that the first two of these equations define the induced world-sheet torsion components $T^{(\pm2)}\equiv de^{(\pm2)}\pm e^{(\pm2)}\wedge(\omega+\bar\omega)(d)$ via the wedge product of the covariant differentials for the Grassmann-odd components of supertwistors. The last equation from the set (\ref{55}) together with the nontrivial relation that stems from the integrability conditions for the admissible supertwistor differentials (\ref{33}) yields that
\begin{equation}
e^{(+2)}\wedge\tilde\omega(d)+{\textstyle\frac{i(s+1)}{2c(\alpha^\prime)^{1/2}}}{\cal D}\bar\eta\wedge{\cal D}\zeta=0,\quad e^{(-2)}\wedge\Omega(d)-{\textstyle\frac{i(s-1)}{2c(\alpha^\prime)^{1/2}}}{\cal D}\bar\eta\wedge{\cal D}\zeta=0.
\end{equation}

Finally the fermionic equations of motion have the form
\begin{equation}
\begin{array}{c}
\frac{\delta S^{N=1}}{\delta{\cal D}\bar\eta}=e^{(+2)}\wedge{\cal D}\eta+\frac{is}{2c(\alpha^\prime)^{1/2}}[\omega_{\cal W}(d)\wedge{\cal D}\eta-\omega_{{\cal W}\bar{\cal Z}}(d)\wedge{\cal D}\zeta]=0,\\[0.2cm]
\frac{\delta S^{N=1}}{\delta{\cal D}\bar\zeta}=e^{(-2)}\wedge{\cal D}\zeta-\frac{is}{2c(\alpha^\prime)^{1/2}}[\omega_{\cal Z}(d)\wedge{\cal D}\zeta-\omega_{{\cal Z}\bar{\cal W}}(d)\wedge{\cal D}\eta]=0.
\end{array}
\end{equation}
The substitution of (\ref{srheo}), (\ref{srheo2}) transforms them to the following equations
\begin{equation}\label{57}
(s+1)e^{(+2)}\wedge{\cal D}\eta=0,\quad (s-1)e^{(-2)}\wedge{\cal D}\zeta=0.
\end{equation}
One observes that the definite choice of the numerical coefficient $s=\pm1$ turns one of the above equations into identity implicitly exhibiting the $\kappa$-invariance of the action functional (\ref{sust}) in the supertwistor representation.

For the completeness of presentation let us adduce the $\kappa$-symmetry transformations for the supertwistor formulation of the $N=1$ superstring action. When $s=-1$ one has
\begin{equation}
\begin{array}{c}
\delta_\kappa{\cal Z}^A=\tilde\omega(\delta_\kappa){\cal W}^A-\zeta\kappa{\cal I}^{AB}\bar{\cal Z}_B+\eta\kappa{\cal I}^{AB}\bar{\cal W}_B+{\cal J}^{A}_{B}{\cal K}^B,\ \delta_\kappa{\cal W}^A=\Omega(\delta_\kappa){\cal Z}^A,\\[0.2cm]
\delta_\kappa e_\mu^{(+2)}=0,\quad\delta_\kappa e^{(-2)}_\mu=-\frac{i}{c(\alpha^\prime)^{1/2}}(\kappa{\cal D}_\mu\eta-{\cal D}_\mu\bar\eta\bar\kappa)
\end{array}
\end{equation}
and c.c. expressions, where $\kappa(\xi^\mu)$ is a complex Grassmann-odd transformation parameter written in the supertwistor form as ${\cal K}^A\equiv(0, 0, \kappa)$ and
\begin{equation}
\omega(\delta_\kappa)=0,\ \tilde\omega(\delta_\kappa)={\textstyle\frac{i}{2c(\alpha^\prime)^{1/2}}}e^{(-2)\mu}{\cal D}_\mu\zeta\kappa,\ \bar\Omega(\delta_\kappa)=-{\textstyle\frac{i}{2c(\alpha^\prime)^{1/2}}}e^{(+2)\mu}{\cal D}_\mu\zeta\kappa.
\end{equation}
Corresponding expressions for $s=1$ read
\begin{equation}
\begin{array}{c}
\delta_\kappa{\cal Z}^A=\tilde\omega(\delta_\kappa){\cal W}^A,\ \delta_\kappa{\cal W}^A=\Omega(\delta_\kappa){\cal Z}^A-\zeta\kappa{\cal I}^{AB}\bar{\cal Z}_B+\eta\kappa{\cal I}^{AB}\bar{\cal W}_B+{\cal J}^{A}_{B}{\cal K}^B,\\[0.2cm]
\delta_\kappa e_\mu^{(-2)}=0,\quad\delta_\kappa e^{(+2)}_\mu=-\frac{i}{c(\alpha^\prime)^{1/2}}(\kappa{\cal D}_\mu\zeta-{\cal D}_\mu\bar\zeta\bar\kappa)
\end{array}
\end{equation}
with 
\begin{equation}
\omega(\delta_\kappa)=0,\ \tilde\omega(\delta_\kappa)=-{\textstyle\frac{i}{2c(\alpha^\prime)^{1/2}}}e^{(-2)\mu}{\cal D}_\mu\bar\eta\bar\kappa,\ \bar\Omega(\delta_\kappa)={\textstyle\frac{i}{2c(\alpha^\prime)^{1/2}}}e^{(+2)\mu}{\cal D}_\mu\bar\eta\bar\kappa.
\end{equation}

So it is possible to pose a question whether the tensile $N=1$ superstring action can be simplified similarly to superparticle and tensionless superstring models, where the transition to the supertwistor formulation allows to remove pure gauge degrees of freedom without covariance breaking and explicit gauge fixation? In the remaining part of the paper we try to answer this question. First note that one can substitute rheotropic relations (\ref{srheo}), (\ref{srheo2}) into the Wess-Zumino term of the superstring action (\ref{41}) and sum up the result with the kinetic term (\ref{39}) to obtain
\begin{equation}\label{62}
S|^{N=1}_{r.s.}=\int\limits_{{\rm M}^2}\left({\textstyle\frac{i}{4(\alpha^\prime)^{1/2}}}[e^{(+2)}\wedge(\omega_{\cal Z}+s\omega_\eta)(d)-e^{(-2)}\wedge(\omega_{\cal W}-s\omega_\zeta)(d)]+{\textstyle\frac{c}{2}}e^{(-2)}\wedge e^{(+2)}\right),
\end{equation}
where r.s. is the abbreviation for "rheotropic shell".
In the space-time formulation 1-forms $\omega_{\cal Z}+s\omega_\eta$ and $\omega_{\cal W}-s\omega_\zeta$ are presented as
\begin{equation}
\begin{array}{c}
(\omega_{\cal Z}+s\omega_\eta)(d)=-2i(\bar udxu+2i(1+s)(ud\theta)(\bar u\bar\theta)-2i(1+s)(u\theta)(\bar ud\bar\theta)),\\[0.2cm]
(\omega_{\cal W}-s\omega_\zeta)(d)=-2i(\bar vdxv+2i(1-s)(vd\theta)(\bar v\bar\theta)-2i(1-s)(v\theta)(\bar vd\bar\theta)).
\end{array}
\end{equation}
So that for any of the admissible values of $s=\pm1$ one of the above 1-forms becomes purely bosonic. For the sake of definiteness consider below the case $s=+1$. Then the action (\ref{62}) acquires the form
\begin{equation}\label{66}
S|^{N=1}_{r.s.}\!=\!\int\limits_{{\rm M}^2}\!\left({\textstyle\frac{i}{4(\alpha^\prime)^{1/2}}}[e^{(+2)}\!\wedge\!(\tilde{\cal Z}^Ad\bar{\tilde{\cal Z}}_A-d\tilde{\cal Z}^A\bar{\tilde{\cal Z}}_A)-e^{(-2)}\!\wedge\!(W^ad\bar W_a-dW^a\bar W_a)]+{\textstyle\frac{c}{2}}e^{(-2)}\!\wedge e^{(+2)}\right),
\end{equation}
where
\begin{equation}\label{67}
\begin{array}{c}
\tilde{\cal Z}^A=(\tilde\mu^\alpha, \bar u_{\dot\alpha}, \bar{\tilde\eta}):\quad\tilde\mu^\alpha=i\bar u_{\dot\alpha}x^{\dot\alpha\alpha}+\theta^\alpha\bar{\tilde\eta},\quad\bar{\tilde\eta}=4\bar u^{\dot\alpha}\bar\theta_{\dot\alpha};\\[0.2cm]
\bar{\tilde{\cal Z}}_A=(\bar u_{\alpha}, \bar{\tilde\mu}^{\dot\alpha}, \tilde\eta):\quad\bar{\tilde\mu}^{\dot\alpha}=-ix^{\dot\alpha\alpha}u_\alpha-\bar\theta^{\dot\alpha}\tilde\eta,\quad\tilde\eta=4u^\alpha\theta_\alpha;\\[0.2cm]
\tilde{\cal Z}^A\bar{\tilde{\cal Z}}_A=\tilde\mu^\alpha u_\alpha+\bar u_{\dot\alpha}\bar{\tilde\mu}^{\dot\alpha}-\frac12\bar{\tilde\eta}\tilde\eta=0
\end{array}
\end{equation}
and $W^a$ and $\bar W_a$ are bosonic twistors. For $s=-1$ the role of $Z$ and $W$ twistors is interchanged. We observe that since the action (\ref{66}) contains only 2 of 4 space-time Grassmann variables, going to the rheotropic shell in the supertwistor formulation of the $N=1$ superstring action allows to exclude 2 pure gauge components of the superspace coordinates $\theta^\alpha$, $\bar\theta^{\dot\alpha}$ without violation of the Lorentz covariance. Note that the obtained action (\ref{66}) is that of the $2d$ free field theory of the heterotic type and comprises only physical degrees of freedom in the fermionic sector.

Consider a generalization of the above results to the $N=2$ superstring case. It turns out that because of the presence of the 4-fermion contribution to the Wess-Zumino term it is not enough to use the rheotropic relations to bring the action to the quadratic form similar to (\ref{66}) it is necessary to explicitly fix the $\kappa$-symmetry gauge. Indeed, the $N=2$ superstring action is given by the expression
\begin{equation}\label{n2}
S^{N=2}=S^{N=2}_{kin}+S^{N=2}_{WZ}
\end{equation}
with
\begin{equation}
S^{N=2}_{kin}=\int\limits_{{\rm M}^2}\left({\textstyle\frac{1}{2(\alpha^\prime)^{1/2}}}[e^{(+2)}\wedge(\bar u\omega(d)u)-e^{(-2)}\wedge(\bar v\omega(d)v)]+{\textstyle\frac{c}{2}}e^{(-2)}\wedge e^{(+2)}\right),
\end{equation}
where $\omega_{\alpha\dot\alpha}(d)=dx_{\alpha\dot\alpha}+2id\theta^I_\alpha\bar\theta^I_{\dot\alpha}-2i\theta^I_\alpha d\bar\theta^I_{\dot\alpha}$, $I=1,2$,
and
\begin{equation}
\begin{array}{rl}
S^{N=2}_{WZ}=&\frac{is}{c\alpha^\prime}{\displaystyle\int\limits_{{\rm M}^2}}\omega^{\dot\alpha\alpha}(d)\wedge(d\theta^1_\alpha\bar\theta^1_{\dot\alpha}-\theta^1_\alpha d\bar\theta^1_{\dot\alpha}-d\theta^2_\alpha\bar\theta^2_{\dot\alpha}+\theta^2_\alpha d\bar\theta^2_{\dot\alpha})\\[0.2cm]
+&\frac{2s}{c\alpha^\prime}{\displaystyle\int\limits_{{\rm M}^2}}(d\theta^{1\alpha}\bar\theta^{1\dot\alpha}-\theta^{1\alpha}d\bar\theta^{1\dot\alpha})\wedge(d\theta^2_\alpha\bar\theta^2_{\dot\alpha}-\theta^2_\alpha d\bar\theta^2_{\dot\alpha}),
\end{array}
\end{equation}
where as before $s=\pm1$.
Then we covariantly fix $\kappa$-symmetry gauge by the relation
\begin{equation}
\theta^1_\alpha=\theta^2_\alpha
\end{equation}
that leads to the vanishing of the Wess-Zumino term as the whole. As for the remaining kinetic term, using the relations similar to (\ref{38}) it can be brought to the form
\begin{equation}
\begin{array}{rl}
 S|^{N=2}_{\kappa-gauge}=&{\textstyle\frac{i}{4(\alpha^\prime)^{1/2}}}{\displaystyle\int\limits_{{\rm M}^2}}[e^{(+2)}\wedge(\tilde{\cal Z}^{A}d\bar{\tilde{\cal Z}}_A-d\tilde{\cal Z}^{A}\bar{\tilde{\cal Z}}_A)-e^{(-2)}\wedge(\tilde{\cal W}^{A}d\bar{\tilde{\cal W}}_A-d\tilde{\cal W}^{A}\bar{\tilde{\cal W}}_A)]\\[0.2cm]
+&{\textstyle\frac{c}{2}}{\displaystyle\int\limits_{{\rm M}^2}}e^{(-2)}\wedge e^{(+2)},
\end{array}
\end{equation}
where the supertwistors $\tilde{\cal Z}^A$ and $\bar{\tilde{\cal Z}}_A$ are given by the expressions (\ref{67}) and
\begin{equation}
\begin{array}{c}
\tilde{\cal W}^{A}=(\tilde\nu^{\alpha}, \bar v_{\dot\alpha}, \bar{\tilde\zeta}):\quad\tilde\nu^{\alpha}=i\bar v_{\dot\alpha}x^{\dot\alpha\alpha}+\theta^{\alpha}\bar{\tilde\zeta},\ \bar{\tilde\zeta}=4\bar v^{\dot\alpha}\bar\theta_{\dot\alpha};\\[0.2cm]
\bar{\tilde{\cal W}}_A=(v_\alpha, \bar{\tilde\nu}{}^{\dot\alpha}, \tilde\zeta):\quad\bar{\tilde\nu}{}^{\dot\alpha}=-ix^{\dot\alpha\alpha}v_\alpha-\bar\theta^{\dot\alpha}\tilde\zeta,\ \tilde\zeta=4v^\alpha\theta_\alpha.
\end{array}
\end{equation}

\section{Discussion and conclusion}

In the paper addressed was the issue of the supertwistor
formulation of the superstring action in $D=4$
superspace that is one of the classically allowed dimensions for
the superstring to live in and where the twistor methods are most
efficient and well developed. We were stimulated by the instances
of superparticle, tensionless superstring and super $p$-brane
models, where the transition to the supertwistor representation
results in the covariant reduction of some of the redundant
degrees of freedom including pure gauge Grassmann variables
related to the $\kappa$-symmetry and the simplification of the
covariant quantization procedure for superparticles.

It appears that the most suitable for the transition to the
(super)twistor representation is the Lorentz-harmonic formulation
of the (super)string action classically equivalent to the original
one and characterized by the presence of the auxiliary spinor
variables that in $D=4$ space-time coincide with the
Newman-Penrose dyad components and ensure the covariant solution
of the Virasoro constraints and hence the irreducible realization
of the $\kappa$-symmetry in the supersymmetric model. The
(super)twistor form of the action was obtained by the Penrose
change of variables. In sections 2 and 3 we derived the
(super)string equations of motions in the (super)twistor form that
required working out the proper variational principle taking into
account the (super)twistor constrained nature. Also there were
explicitly presented irreducible $\kappa$-symmetry transformations
in the supertwistor realization exhibiting that unlike the
superparticle, tensionless superstring and super $p$-brane models
the reduction of pure gauge fermionic variables for the tensile
superstring does not occur just by passing to the supertwistor
variables. We showed that it is possible to exclude such
unphysical variables by the implicit $\kappa-$symmetry gauge
fixation via the substitution into the Wess-Zumino term of the
$N=1$ superstring action of the rheotropic relations that are
nondynamical equations of motion following from the kinetic part
of the superstring action upon variation with respect to the
zweibein and dyad components. Thus the $N=1$ superstring action
acquires the manifestly heterotic type form as the sum of the quadratic term
in bosonic twistor variables for the left(right) movers and of the
quadratic term in the supertwistors for the right(left) movers. We
have also established that in order to bring to the similar
quadratic form the $N=2$ superstring action one has to fix
explicitly the $\kappa$-symmetry gauge due to the nonlinear
contribution of the Grassmann variables to the Wess-Zumino term.

In this paper we confined ourselves to the examination of the supertwistor formulation of the superstring in $D=4$ $N=1,2$ superspaces using the Lagrangian approach. The features of Hamiltonian and quantum dynamics of the superstring will be studied in the forthcoming paper. Further lines of the development of presented results include the generalization to other allowed dimensions of the particular interest among which is the $D=10$ superstring, as well as, the exploration of (super)string models in complex (super)spaces and the examination of the supertwistor formulations for super $p$-branes.

\section{Acknowledgements}

It is a pleasure to thank A.A.~Zheltukhin for the valuable discussions. The author is also grateful to the Abdus Salam ICTP, where part of this work was done, for the warm hospitality. The work was partially supported by the Young Researchers' Grant of the NAS of Ukraine.

\end{document}